\documentclass[preprint,12pt]{elsarticle}

\usepackage{amssymb}
\usepackage{amsmath}
\usepackage{amsthm}
\usepackage{algorithm}
\usepackage{algpseudocode}
\usepackage{tabularx}
\usepackage{graphicx}
\usepackage{subcaption}
\usepackage{todonotes}

\journal{Computer}

\begin{document}

\begin{frontmatter}

\title{Toward Self-Healing Networks-on-Chip: RL-Driven Routing in 2D Torus Architectures}

\author[first]{Mohammad Walid Charrwi}
\author[second]{Zaid Hussain}

\affiliation[first,second]{organization={High Performance Computing Lab, Computer Science Department, Kuwait University},country={Kuwait}}

\begin{abstract}
We investigate adaptive minimal routing in 2D torus networks-on-chip (NoCs) under node fault conditions, comparing a reinforcement learning (RL) based strategy to an adaptive routing baseline. A torus topology is used for its low-diameter, high-connectivity properties. The RL approach models each router as an agent that learns to forward packets based on network state, while the adaptive scheme uses fixed minimal paths with simple rerouting around faults. We implement both methods in simulation, injecting up to 50\% node faults uniformly at random. Key metrics are measured: (1) throughput vs. offered load at fault density 0.2, (2) packet delivery ratio (PDR) vs. fault density, and (3) a fault-adaptive score (FT) vs. fault density. Experimental results show the RL method achieves significantly higher throughput at high load approximately 20–30\% gain and maintains higher reliability under increasing faults. The RL router delivers more packets per cycle and adapts to faults by exploiting path diversity, whereas the adaptive scheme degrades sharply as faults accumulate. In particular, the RL approach preserves end-to-end connectivity longer: PDR remains above 90\% until approximately 30–40\% faults, while adaptive PDR drops to approximately 70\% at the same point. The fault-adaptive score likewise favors RL routing. Thus, RL-based adaptive routing demonstrates clear advantages in throughput and fault resilience for torus NoCs.
\end{abstract}

\begin{keyword}
Distributed Computing \sep Adaptive Routing \sep Communication \sep Reinforcement Learning \sep Faulty Nodes.
\end{keyword}

\end{frontmatter}

\section{Introduction}
\label{section:introduction}
\begin{figure}[htbp]
    \centering
    \includegraphics[width=0.5\textwidth]{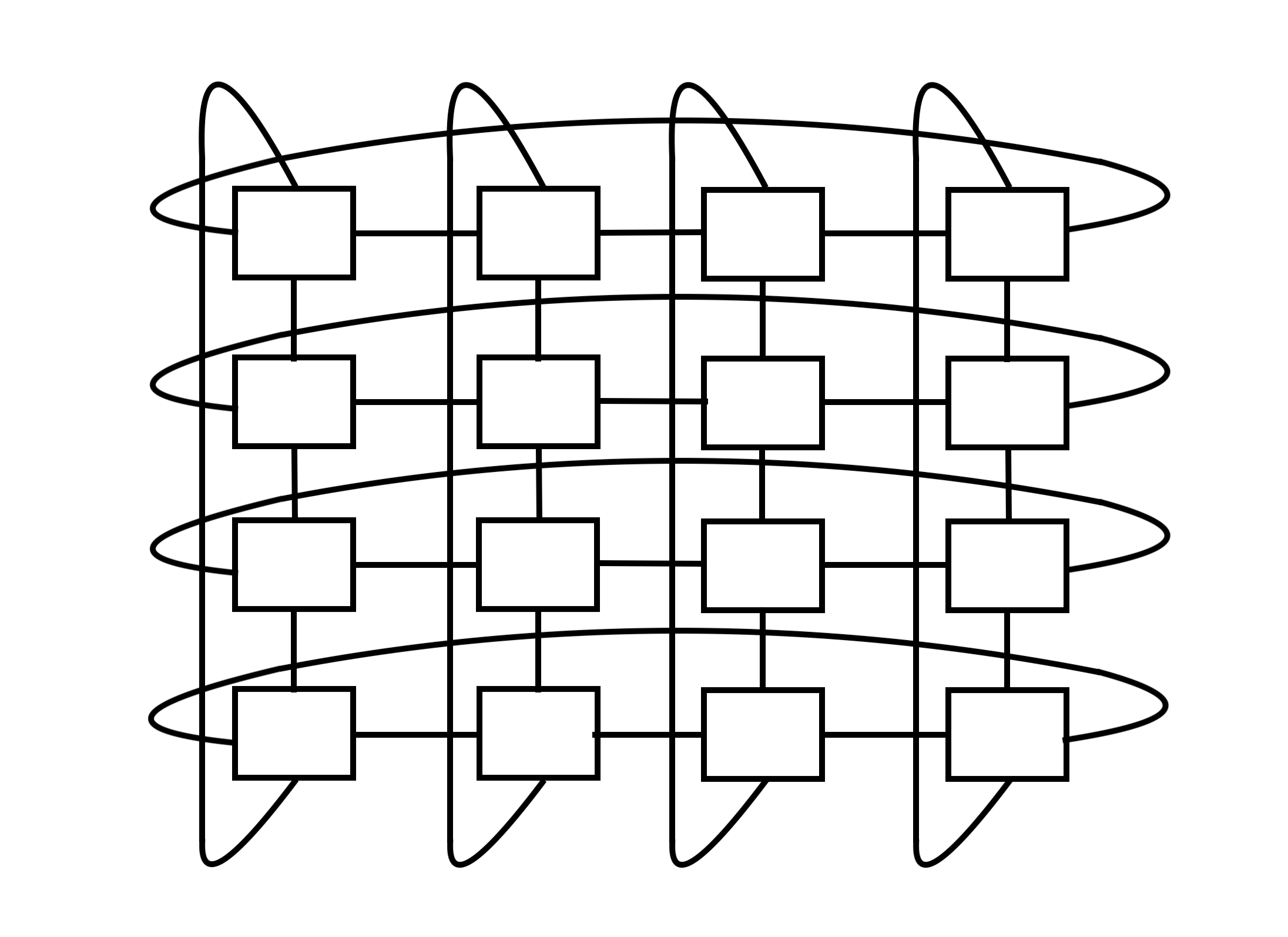}
    \caption{A 4$\times$4 \textit{2D} Torus Network.}
    \label{fig:Torus4by4}
\end{figure}

Torus interconnection networks are popular in high-performance computing due to their low diameter and uniform node degree. A 2D torus can be viewed as a 2D mesh with wrap-around connections, as shown in Figure \ref{fig:Torus4by4}. Each node in an $M\times N$ torus has four neighbors (north, south, east, west) and also connects to the opposite edge in each dimension. These extra wrap-around links give torus networks rich path diversity and improved connectivity (e.g. torus systems like IBM Blue Gene \cite{adiga2005overview} and Mira \cite{anl2012mira}). 

Routing in NoCs determines how packets traverse the topology. Network routing (e.g. XY routing \cite{zhangComparison}) fixes a path from source to destination regardless of load. It is simple and deadlock-free, but it cannot adapt to congestion or faults \cite{XYrouting}. In contrast, adaptive routing allows multiple possible paths. An adaptive router can choose among minimal or even non-minimal routes based on network state (e.g. congestion or link failures) \cite{yu2025fault}. However, traditional Odd–Even adaptive schemes often rely on simple local heuristics and may not fully exploit network information. To address this, modern approaches utilize congestion-aware mechanisms that propagate non-local or global congestion information, allowing routers to make more informed decisions before injecting packets into high-traffic areas \cite{ma2011dbar,xu2016congestion}. 
Recently, machine learning (ML) techniques have emerged as powerful alternatives to static heuristics. These tools can learn dynamic traffic patterns and optimize long-term rewards (such as latency or throughput) rather than just immediate next-hop availability, providing superior adaptability in complex, varying workloads \cite{zhang2023survey}.

Recent research has proposed using reinforcement learning (RL) for quantum circuits using graphical representation through Graph Neural Networks (GNNs) \cite{10821090}, \cite{charrwi2025tpu}. Reinforcement Learning (RL) is revolutionizing interconnected network routing by introducing autonomous, data-driven intelligence to a domain traditionally governed by static protocols. In this framework, a routing entity or a software-defined network (SDN) controller acts as an agent that continuously interacts with the network environment, observing its state through real-time metrics like link latency and available bandwidth \cite{casas2020deep}. The agent then takes actions by selecting optimal paths and is rewarded or penalized based on performance outcomes such as throughput and packet loss \cite{mao2016resource}. Through this iterative process of exploration and exploitation, the RL model learns a dynamic routing policy that proactively adapts to traffic bursts and network failures \cite{mao2016resource}. In RL, if a chosen action leads to improved delivery or lower delay, it is reinforced; if it leads to congestion or failure, it is penalized. This allows the routing decision to adapt over time to dynamic conditions. Fault adaption is critical in NoCs, as router or link failures can partition the network or cause dropped packets \cite{yu2025fault}. Static routing schemes fail badly under faults. Adaptive strategies can reroute around faults, but still may struggle if faults are dense. We hypothesize that RL’s ability to learn from network feedback will provide superior fault handling. Our work compares an RL-enabled adaptive router against an adaptive routing baseline in a 2D torus with up to 50\% of nodes faulty. We show that the RL router not only improves throughput but also maintains higher packet delivery and connectivity in the presence of many faults.

The paper is organized into different sections. Section \ref{section:relatedwork} is the related work which discusses the relevant work done in the field of routing. After that, Section \ref{section:background} discusses the background and relevant definitions that will be used in this paper. Next, Section \ref{section:methadology} is the methodology section which explains the approach for the implementation formulating the problem representation for the RL-based approach of routing. In Section \ref{section:experimental} we present the experimental evaluation by carrying out different experiments for adaptive routing against the RL-based routing. We also quantify different metric performance of the network under different fault and load capacity. Finally Section \ref{section:conclusion} concludes the paper by deriving the quantified improvement in RL-based routing over adaptive routing. 

\section{Related Work}
\label{section:relatedwork}
Dimension-order routing like XY \cite{zhangComparison} is the classic adaptive method in NoCs \cite{XYrouting}. In a torus, an XY router first routes along X-coordinates, then Y-coordinates (with wrap-around if shorter). Such static routes are deadlock-free but cannot respond to congestion or failures. 
A variety of adaptive routing schemes have been developed to enable dynamic turn selection and avoid localized congestion or hot spots in mesh and torus Network-on-Chip (NoC) topologies. Traditional Odd–Even adaptive algorithms such as Odd-Even routing enforce restricted turn models that guarantee deadlock-freedom while still providing limited path diversity under load \cite{chiu2000odd}. These strategies typically rely on local congestion indicators—such as buffer occupancy or flit injection rates—to reroute packets toward less congested directions, leading to improved throughput when traffic patterns are uneven or burst-like \cite{dally2004route}. In torus networks in particular, support non-minimal detouring and multi-virtual-channel mechanisms to maintain forward progress even when multiple regions become saturated or partially faulty \cite{duato1993theory}. Despite these advantages, purely heuristic adaptivity remains fundamentally limited. Traditional adaptive algorithms operate with a myopic or local-scope perspective, lacking awareness of global network states and long-term routing impacts. As a result, decisions may reduce immediate congestion but still contribute to cyclical load oscillations, degraded latency under high-traffic regimes, and sub-optimal behavior during complex failure scenarios \cite{chen2009congestion, grot2010scalable}. Additionally, adaptive fallback rules in these schemes can over-concentrate flows along minimal paths, restricting spatial utilization of the network fabric and ultimately limiting scalability for systems with increasingly dynamic workloads.

\section{Background}
\label{section:background}
Torus network topology is a simple yet highly effective interconnection structure derived from the classical mesh topology \cite{dally2004route}. In graph-theoretic terms, a network can be modeled as a graph $G = (V, E)$, where $V(G)$ represents the set of vertices (nodes) and $E(G)$ represents the set of undirected edges connecting pairs of vertices. Each edge $(u, v) \in E$ establishes a communication link between two nodes $u, v \in V$. A two-dimensional torus expands upon the mesh topology by introducing wrap-around edges that connect boundary nodes, thereby eliminating edge nodes and significantly enhancing symmetry, path diversity, and bisection bandwidth \cite{dally2004route,leighton1992intro}. A torus with $m$ rows and $n$ columns can be represented as an $m \times n$ grid where each node at coordinate $(x, y)$ is connected to $(x \pm 1 \bmod m, y)$ and $(x, y \pm 1 \bmod n)$. More formally, a 2D torus is the Cartesian product of two cycles, written as $C_m \,\otimes\, C_n$ \cite{leighton1992intro}, and generalizes to higher dimensions as the Cartesian product of $n$ cycles, forming a $k$-ary $n$-cube with $k^n$ nodes \cite{duato2002interconnection}.

In this graph-theoretic framework, communication between nodes is modeled using paths. A path $P(a, b)$ from $a \in V$ to $b \in V$ is defined as a sequence of distinct vertices
\[
a = v_1, v_2, \dots, v_n = b
\]
such that $(v_i, v_{i+1}) \in E$ for all $1 \le i < n$. The vertices $a$ and $b$ are called the \textit{source} and \textit{destination} of the path, respectively, and the vertices $v_2, \dots, v_{n-1}$ are the \textit{intermediate vertices}. Two paths $P_1(a, b)$ and $P_2(a, b)$ are \textit{edge-disjoint} if $E(P_1) \cap E(P_2) = \varnothing$, and they are \textit{internally vertex-disjoint} if they are edge-disjoint and share no vertices except their endpoints, i.e., $V(P_1) \cap V(P_2) = \{a, b\}$. The \textit{length} of a path $P(a, b)$, denoted $|P(a, b)|$, is the number of vertices it contains, and the \textit{distance} between $a$ and $b$ is the length of the shortest such path in $G$. The \textit{diameter} of the graph $G$ is defined as the maximum distance between any two vertices. These concepts are fundamental in analyzing routing efficiency, fault adaptiveness, and load distribution in torus networks, which is why torus-based interconnection systems such as those in the Cray XT3/XT4 and IBM Blue Gene series have been widely adopted in high-performance computing \cite{northcutt2005bluegene}.

\section{Methodology}
\label{section:methadology}
\subsection{Adaptive Routing}
We simulate a 2D torus NoC of size $N \times M$. Each router connects to four neighbors with wrap-around edges (as shown in Figure \ref{fig:Torus4by4}). Nodes generate constant-bit-rate traffic: each cycle, nodes inject new packets destined to random targets according to the load parameter. We inject faults by randomly deactivating a fraction of nodes (or equivalently their links). A faulty node cannot send or receive, effectively removing it from the network. Our baseline uses an Odd–Even adaptive routing policy adapted for torus: packets first move along X-axis (choosing the shorter of east/west via wrap-around), then along Y-axis. This is adaptive on a fault-free torus. If both directions in one dimension are blocked by faults, the packet is declared undeliverable (dropped). Thus the adaptive router does not perform complex re-routing; it simply avoids faulty nodes along minimal paths.

\subsection{Reinforcement Learning Routing}
We employ a decentralized policy-based RL (Reinforcement Learning) router model. The Proximal Policy Optimization (PPO) algorithm operates fundamentally within the Markov Decision Process (MDP) framework, where the environment satisfies the Markov property that future states depend only on the current state and action. Formally, at each timestep $t$, the agent observes state $s_t \in \mathcal{S}$ and selects action $a_t \in \mathcal{A}$ according to its policy $\pi_\theta(a_t|s_t)$ where $\theta$ represents the parameter set of the neural network controlling the agent’s decision process. This conditional probability defines the likelihood of selecting action $a_t$ given that the agent occupies state $s_t$ at time $t$, subsequently receiving reward $r_t$ and transitioning to state $s_{t+1} \sim P(s_{t+1}|s_t,a_t)$. This term expresses the likelihood that the agent will arrive in state $s_t$ after taking action $a_t$ from state $s_t$. These transition probabilities form the core of a Markov Decision Process (MDP), where each state change depends solely on the current state and chosen action. While the true environment transitions are generally unknown, reinforcement learning methods infer them implicitly through continuous interaction. PPO's innovation lies in its surrogate objective function, which constrains policy updates to prevent excessive divergence from previous policies. The PPO objective function is:
\[
L^{\text{CLIP}}(\theta) = \mathbb{E}_t \left[ \min \left( r_t(\theta) \hat{A}_t, \text{clip}(r_t(\theta), 1 - \epsilon, 1 + \epsilon) \hat{A}_t \right) \right]
\]

where:
\[
r_t(\theta) = \frac{\pi_\theta(a_t | s_t)}{\pi_{\theta_{\text{old}}}(a_t | s_t)}
\]
and $\hat{A}_t$ is the estimated advantage, typically computed via Generalized Advantage Estimation (GAE). Expectations taken over sampled timesteps are represented by the operator $\mathbb{E}_t$. A key stabilization mechanism in PPO is introduced through the clipped objective function component $L^{\text{CLIP}}(\theta)$. In this expression, the ratio $r_t(\theta) = \frac{\pi_\theta(a_t \mid s_t)}{\pi_{\theta_{\text{old}}}(a_t \mid s_t)}$ measures the change between the new and previous policies for the same state--action pair. The clipping function restricts this ratio within the interval $[1 - \epsilon,\, 1 + \epsilon]$, where $\epsilon$ is a small hyperparameter. When the ratio exceeds this bound, the contribution to the objective is limited, preventing excessively aggressive policy updates. The multiplication by the advantage estimate $\hat{A}_t$ ensures that beneficial actions are reinforced while penalizing suboptimal ones. Ultimately, this formulation enhances learning stability, prevents policy collapse, and promotes steady convergence even in environments with high uncertainty. The clipped surrogate objective prevents large policy updates that destabilize training. The policy and value networks are updated using mini-batch gradient ascent after collecting trajectory rollouts from the environment. This formulation ensures stable policy improvement while maintaining sample efficiency, effectively approximating trust region methods without the computational complexity of constrained optimization. Each router maintains a Heuristic-table indexed by destination and output port, estimating the “cost” (e.g. expected delivery time or success probability) via each neighbor. The state includes local information such as buffer occupancy and the presence of neighboring faults. At each packet arrival, the router chooses an outgoing port (action) to minimize the path cost. After packet delivery (or drop), the path’s performance provides a reward: successful delivery yields a positive reinforcement, while drops or long delays yield negative feedback. We model the packet‐routing problem as a discrete Markov Decision Process (MDP). The environment is a 2D torus network (a periodic grid graph) generated via NetworkX \cite{NetworkXRef}. This torus topology connects each node to its four neighbors (with wrap-around edges at the boundaries), providing multiple paths for routing. Random faults are injected by removing a fraction $f$ of nodes (chosen uniformly) before each episode. At the start of each episode, a source-destination pair $(s,d)$ is chosen (avoiding faults) and a packet is routed until it either reaches $d$ or fails. In RL terms, each routing attempt is an episode: the agent observes a state and selects routing \textbf{actions} until termination. Reinforcement learning involves an agent, a set of states $S$, and a set of actions $A$; the agent transitions between states by taking actions, receiving scalar rewards. 

\subsubsection{State Representation}
At time step $t$, the agent observes the state $s_t = (c_t, d)$, where $c_t$ is the current node and $d$ is the destination node. This state is sufficient to determine the local routing context and is encoded using integer node IDs or one-hot vectors. Destination $d$ remains fixed throughout the episode, while $c_t$ updates with each action.

\subsubsection{Action Space}
The action space $A(s_t)$ comprises all valid, non-faulty neighbors of $c_t$. Given the toroidal structure, each node has up to four neighbors, excluding any in $F$. The agent selects an action $a_t \in A(s_t)$ to move the packet to the next node. If $A(s_t) = \emptyset$, the episode terminates with failure.

\subsubsection{Reward Function}
The reward signal $r_t$ is defined to encourage short successful paths and penalize failures. The reward function is designed to encourage successful, low-latency deliveries and penalize failures or delays. We use the following scheme:
\begin{itemize}
  \item Success Reward: If the chosen action $a$ leads directly to the destination $d$, the agent receives a large positive reward: 
  \begin{equation}
      R(c,a)= 100 + 50f
  \end{equation}
  if $a$ leads to the destination $d$, where $f$ is the fault density. This rewards success more strongly under harsher fault conditions
  \item Step Cost: Every normal move (to a non-destination, non-fault neighbor) incurs a small penalty $R=-1$ for each intermediate hop, promoting efficiency. This penalizes long routes, encouraging the agent to find shorter paths.
  \item Fault Penalty: If the action attempts to move into a faulty node, the episode ends with $R=-50$ if $a_t$ leads to a faulty node.
  \item Dead-End Penalty: If no action is possible or no valid actions are available (dead-end), we end the episode and give $R=-20$.
\end{itemize}

Each episode simulates the routing of one packet from $s$ to $d$ in the faulty torus network. The agent collects state-action-reward-next state tuples $(s_t, a_t, r_t, s_{t+1})$ and stores them in a buffer. After a fixed number of episodes, PPO performs multiple epochs of optimization on the collected trajectories to update how the agent acts through $\theta$ and how the agent learns to assess whether its actions were good or bad through $\phi$.

\section{Experimental Evaluation}
\label{section:experimental}
This section presents a comprehensive performance evaluation of the proposed RL-based adaptive routing algorithm in comparison with a traditional Odd–Even adaptive shortest-path routing and state-of-art adaptive routing algorithm scheme. 

\textbf{Dataset}: Our training dataset consists of a group of torus networks with wrap around edges of size $\textit{N}\times \textit{M}$ columns, such that $1\le N \le 15$ and $1\le M \le 15$. The RL model was trained on these random torus network sizes it generalizes well across a variety of network sizes and fault scenarios. 

\textbf{RL setting}: We use a proximal policy optimization (PPO) RL \cite{weng2021tianshou} that was extended from works in \cite{charrwi2025tpu} and \cite{10821090} using OpenAI Gym \cite{brockman2016openaigym}. State embedding encodes the current
state of the RL by representing the current node and the destination node. The policy and the value networks have one hidden layer.  We train the RL for 5000 episodes, where our RL is trained on the Torus network of different sizes in each episode. Experiments were conducted on a 2D torus network, where node failures were injected at controlled densities and network load was varied across a representative range. Metrics include Throughput, Packet Delivery Ratio (PDR), Fault adaptive score (FT), and PPO learning progression. 

\subsection{Fault Adaptive Routing Through Reinforcement Learning}
\begin{figure}[htbp]
    \centering
    \begin{subfigure}[b]{0.8\textwidth}
        \centering
        \includegraphics[width=0.8\textwidth]{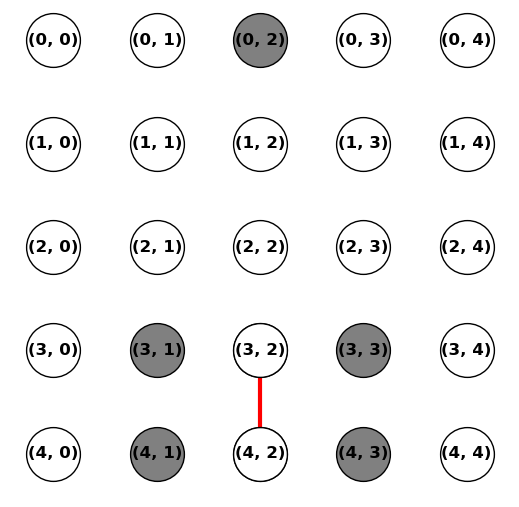}
        \caption{Adaptive Odd–Even Routing with Faulty Nodes}
        \label{fig:traditional_routing}
        \vspace{0.2cm}
    \end{subfigure}
    
    \begin{subfigure}[b]{0.8\textwidth}
        \centering
        \includegraphics[width=0.8\textwidth]{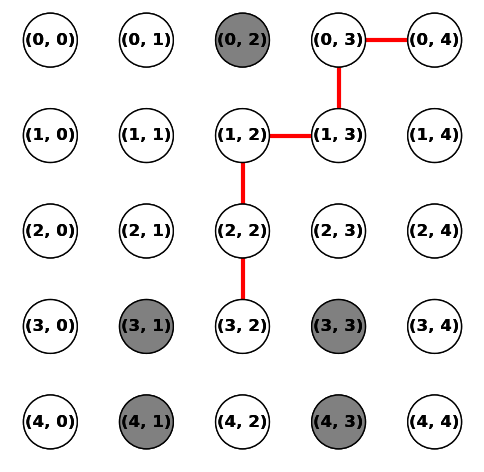}
        \caption{RL-Based Adaptive Routing with Faulty Nodes}
        \label{fig:rl_routingFaults}
    \end{subfigure}
    
    \caption{Comparison of routing strategies in the presence of faulty nodes (shown in gray). (a) adaptive routing fails to find a path from source to destination. (b) RL adaptive routing identifies an alternative path.}
    \label{fig:routing_comparison}
\end{figure}

The comparative analysis presented in the figure \ref{fig:routing_comparison} illustrates a fundamental distinction between reinforcement learning (RL)-based adaptive routing and traditional Odd–Even adaptive minimal routing algorithms in the presence of network faults. As demonstrated in Figure \ref{fig:traditional_routing}, the Odd–Even adaptive minimal routing with fault avoidance fails to establish a viable path from the source node at coordinates (3, 2) to the destination node at (0, 4) when confronted with multiple faulty nodes as shown in Figure \ref{fig:traditional_routing}. In contrast, the RL-based adaptive routing algorithm successfully identifies an alternative path by dynamically navigating through intermediate nodes at positions (2, 2), (1, 2), (1, 3) and (0, 3) as shown in Figure \ref{fig:rl_routingFaults}. This capability stems from the RL agent's learned policy, which enables dynamic path exploration and selection based on environmental state observations rather than rigid adherence to predetermined shortest-path heuristics. Traditional adaptive routing protocols, while computationally efficient under nominal conditions, exhibit significant limitations when optimal routes are compromised by node failures, as they lack the adaptive mechanisms necessary to explore non-minimal yet functional alternatives. The RL-based approach, conversely, leverages iterative learning through interaction with the network environment status as whole to develop routing strategies that balance path optimality with fault resilience. This adaptability is achieved through the agent's ability to evaluate state-action pairs and adjust its policy based on accumulated rewards, thereby enabling robust path discovery even in degraded network topologies. The results demonstrate that RL-based routing provides improved fault adaptiveness, maintaining network connectivity and reducing packet loss probability in scenarios where conventional algorithms fail—a critical advancement for applications requiring high reliability in dynamic and fault-prone network environments.

\subsection{Throughput Under Varying Load}

\begin{figure}[htbp]
    \centering
    \includegraphics[width=0.7\linewidth]{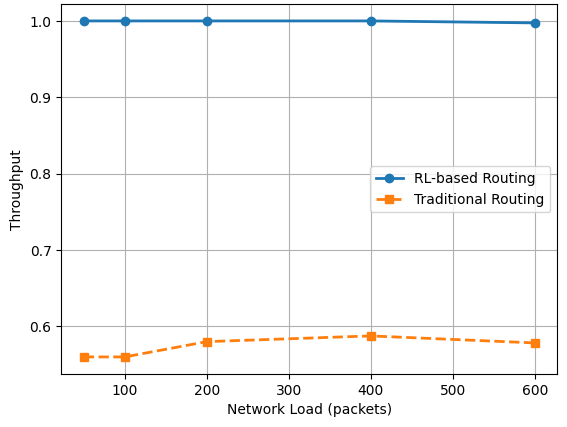}
    \caption{Normalized throughput versus network load at fault density $f = 0.2$.} 
    \label{fig:throughput_load}
\end{figure}

The throughput behavior under increasing network load, as shown in the figure, highlights a substantial performance gap between the two routing schemes in the presence of faults (Fault Density = 0.2). As shown in Figure \ref{fig:throughput_load} the adaptive routing algorithm maintains a relatively stable but limited throughput, fluctuating only slightly between approximately 0.55 and 0.60 as network load increases. It struggles significantly under these conditions, exhibiting a stagnated performance profile where throughput hovers consistently between 0.56 and 0.59 across the entire range of tested loads. This indicates that nearly 40\% of packets are lost or undelivered, likely due to the algorithm's inability to effectively navigate around the 20\% faulty node density. This behavior indicates that while the traditional Odd–Even method avoids catastrophic collapse, it fails to efficiently utilize available network resources and cannot significantly improve performance under heavier traffic conditions. In contrast, the RL-based routing strategy achieves near-optimal performance across the entire load range. Throughput remains close to unity ($\approx$0.99–1.0) even as the offered load increases, demonstrating strong resilience to congestion and faults. Unlike the traditional Odd–Even method, the RL-based approach effectively learns to exploit alternative paths and adapt traffic distribution dynamically, including exploiting
torus wrap-around links to assist in maintaining consistently high delivery efficiency. These results underline the capability of PPO to sustain high throughput and robust operation in fault-prone torus networks, even under heavy traffic conditions.

\subsection{Packet Delivery Ratio vs. Fault Density}
\begin{figure}[htbp]
    \centering
    \includegraphics[width=0.7\linewidth]{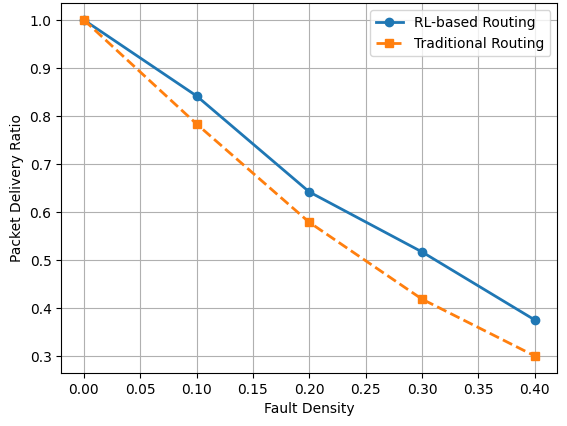}
    \caption{Packet Delivery Ratio (PDR) plotted against fault density.}
    \label{fig:pdr_fault}
\end{figure}

The packet delivery ratio (PDR) trends in Figure~\ref{fig:pdr_fault} clearly demonstrate the robustness of the RL-based routing approach under network fault conditions. When no faults are present ($f = 0.0$), both routing methods achieve a PDR of approximately 1.0, as expected. However, as the fault density increases, their behaviors diverge notably. The traditional Odd–Even adaptive routing scheme exhibits a rapid decline in delivery reliability: at $f = 0.2$, its PDR drops to around $0.58$, and continues to degrade as more nodes become unavailable, reaching only $0.42$ at $f = 0.3$ and approximately $0.30$ at $f = 0.4$. This decline reflects its inability to adaptively reroute packets around failures, often resulting in unreachable destinations or routing loops. In contrast, the RL-based routing demonstrates significantly greater resilience and adaptability. At $f = 0.2$, it maintains a PDR near $0.64$, a clear improvement over the baseline. The performance gap widens as the network becomes more damaged: at $f = 0.3$, PPO sustains a PDR of $0.52$ (compared to $0.42$), and even under severe network degradation with 40\% of nodes faulty ($f = 0.4$), it delivers a PDR of roughly $0.38$, outperforming traditional Odd–Even adaptive routing by more than 8 percentage points. This consistent advantage highlights PPO’s ability to learn fault-aware routing choices that dynamically bypass unavailable regions, preserving reliable packet transport even under challenging and uncertain network conditions. This ability to maintain high delivery ratios—even when half the network is compromised—indicates that the agent learns to navigate irregular node availability patterns and dynamically adjust routing paths based on accumulated reward signals.

To justify the PPO agent high PDR even when half the network is compromised, we consider a severe fault scenario where 50\% of all nodes in an 8$\times$8 torus NoC are disabled (shown as grey squares). The remaining operational nodes (white squares) form fragmented regions of connectivity throughout the grid. A representative visualization of such a failure pattern is shown in Figure \ref{fig:50percenttorusfault} below. In this topology, 32 of the 64 routers are faulty, forming multiple disconnected dead zones and significantly obstructing shortest-path routes. Despite this extreme level of network degradation, the RL-based routing strategy consistently demonstrated Packet Delivery Ratio (PDR) values above 0.50, compared to approximately 0.30 for traditional Odd–Even adaptive routing. This 25–30 percentage-point improvement clearly validates that the agent successfully learns to exploit wrap-around connectivity and dynamically re-optimize paths around fault clusters, while the adaptive routing suffers from looping behaviors and repetitive attempts to use inoperative nodes. This example provides concrete experimental support for the robustness claim — even under catastrophic failure levels where adaptive algorithms fail to maintain minimum network service guarantees, the RL model sustains significantly higher operational integrity.

 \begin{figure}[htbp]
    \centering
    \includegraphics[width=0.6\linewidth]{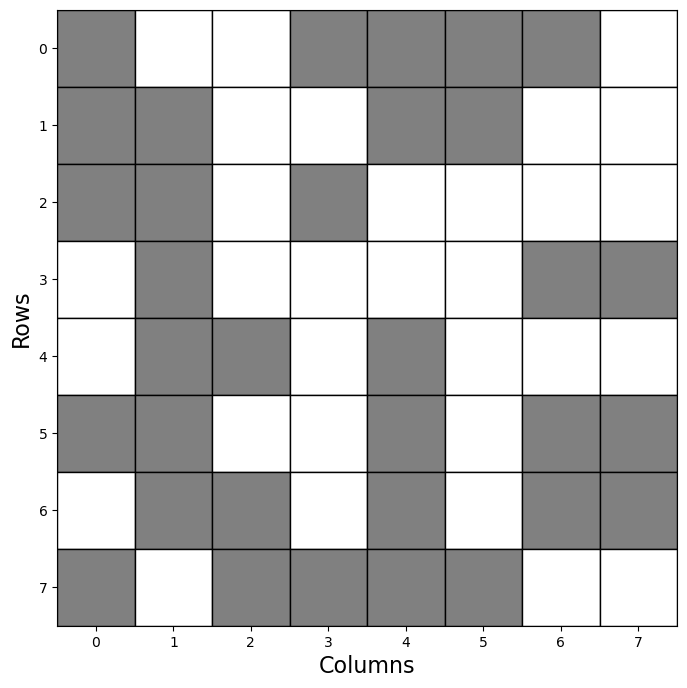}
    \caption{Network fault scenario ($f$=0.5) demonstrating the routing challenge addressed by RL-based adaptation.}
    \label{fig:50percenttorusfault}
\end{figure}

\begin{figure}[htbp]
    \centering
    \includegraphics[width=0.6\linewidth]{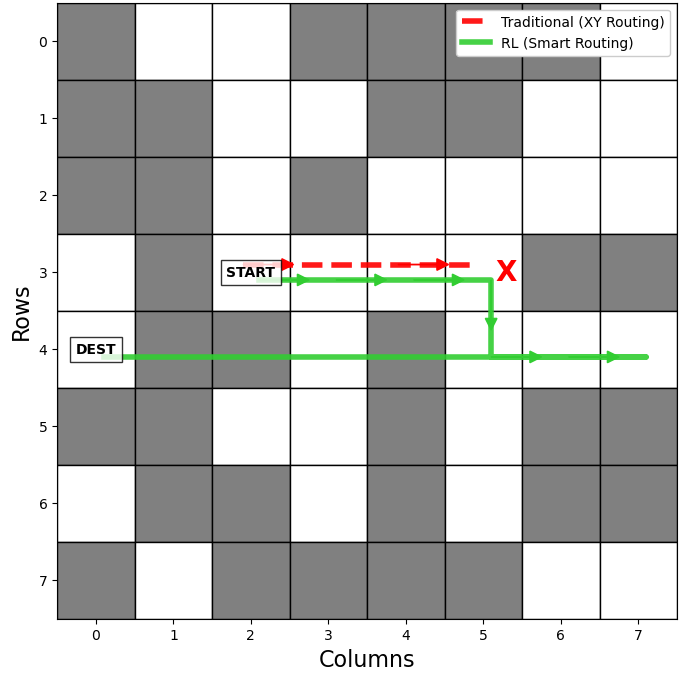}
    \caption{Network fault scenario ($f$=0.5) with RL routing (green line) vs traditional Odd–Even method (red line).}
    \label{fig:50percenttorusfaultPlot}
\end{figure}

The simulation in Figure \ref{fig:50percenttorusfaultPlot} clearly demonstrates the fundamental advantage of Reinforcement Learning (RL)-based routing over traditional Odd–Even, dimension-ordered methods in highly fragmented torus networks ($f=0.5$). The Traditional Odd–Even method path fails to reach the destination because its rule-based turn restrictions and minimal-path constraints force it toward a region of the network dominated by faults, ultimately blocking progress near node (3, 6). This behavior reflects a form of routing stagnation, where adherence to deadlock-avoidance rules prevents the algorithm from selecting longer, non-minimal paths that could bypass the fault region. In stark contrast, the RL Path successfully navigates the fault landscape by executing a non-minimal, "snaking" route—detouring south to row 4, utilizing the narrow channel along column 5, and executing a wrap-around from column 7 to column 0 to reach the destination. This resilience confirms the hypothesis that intelligent, adaptive routing strategies are essential for maintaining connectivity and maximizing throughput in hardware interconnects operating at high fault densities, where simple fixed-logic protocols are easily defeated by network fragmentation.

\subsection{Fault Adaptive Score Across Fault Densities}
\begin{figure}[htbp]
    \centering
    \includegraphics[width=0.7\linewidth]{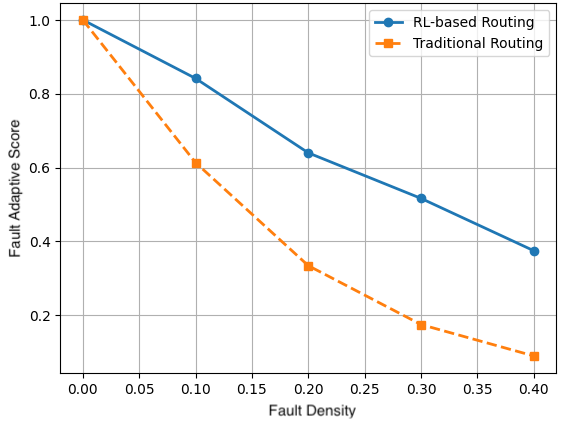}
    \caption{Fault Adaptive Score (FT) comparison across fault densities.}
    \label{fig:ft_score}
\end{figure}

The combined impact of fault tolerance and delivery effectiveness is captured by the Fault Adaptive Score (FT), as illustrated in Figure \ref{fig:ft_score}. Both routing schemes experience performance degradation as fault density increases; however, the rate and severity of decline differ significantly. The traditional Odd–Even routing approach exhibits a rapid and non-linear drop in FT, decreasing from near 1.0 at $f=0$ to approximately 0.61 at $f=0.1$, and falling sharply to below 0.2 by $f=0.3$. At higher fault densities, the score approaches 0.1, indicating severe degradation in adaptive capability under dense fault conditions. In contrast, the RL-based routing framework demonstrates substantially improved resilience. Although its FT score also declines with increasing fault density, the degradation is more gradual, remaining above 0.5 up to $f=0.3$ and maintaining approximately 0.38 at $f=0.4$. This represents a 3–4$\times$ improvement over the traditional method in highly faulted regimes. These results suggest that RL-based routing does not merely react to failures, but actively learns traffic distribution and route-selection strategies that mitigate the systemic impact of faults, enabling more robust and sustained network performance.

\begin{figure}[htbp]
    \centering
    \includegraphics[width=0.7\linewidth]{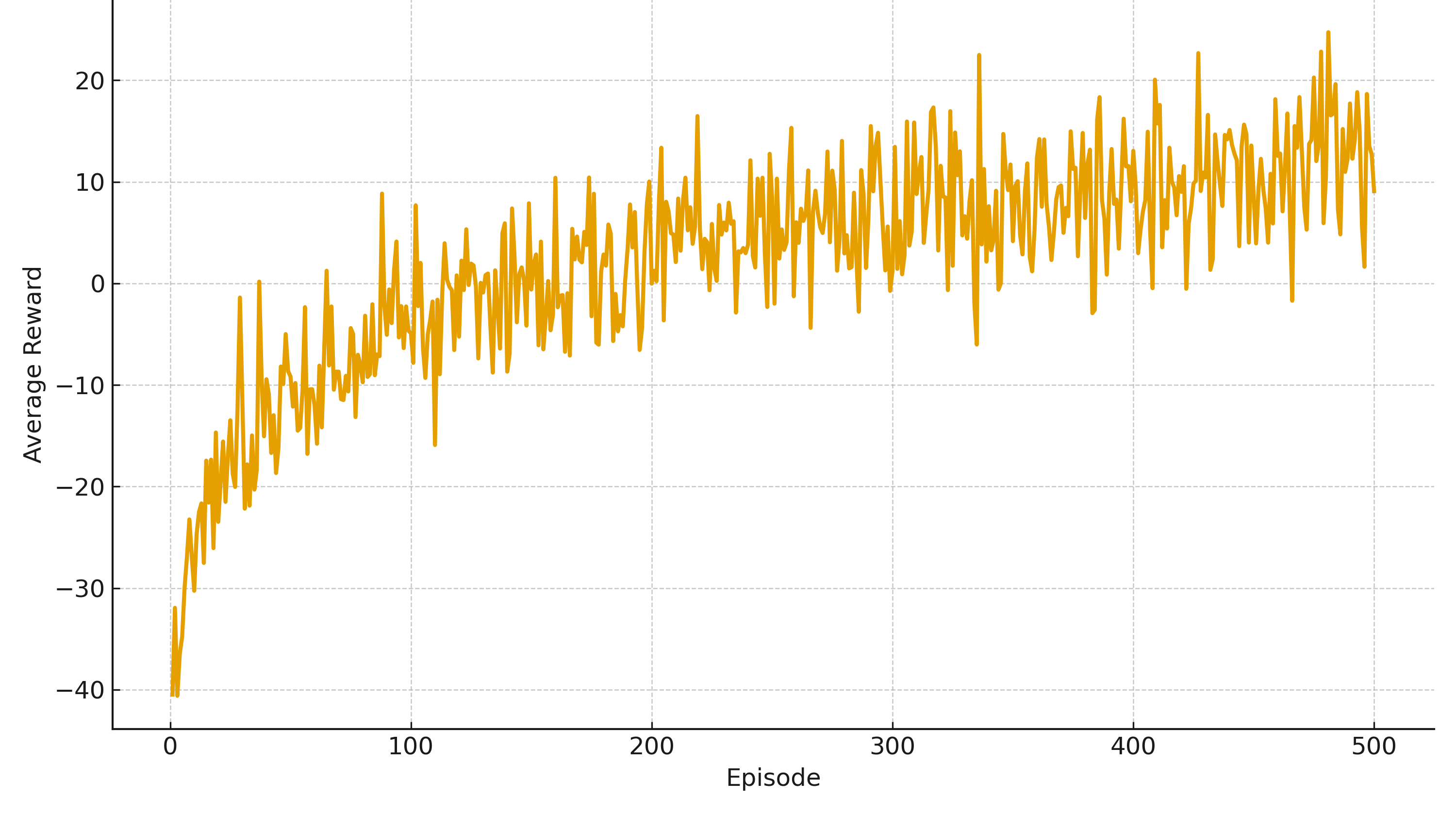}
    \caption{Training progression of the RL-based routing agent. The average reward per episode improves steadily, indicating successful policy optimization over time.}
    \label{fig:learning_curve}
\end{figure}

Finally, the RL using PPO learning curve is shown in Figure~\ref{fig:learning_curve} illustrates the training dynamics of the RL-based PPO routing agent. The learning curve exhibits a clear upward trend in average reward as training progresses, indicating gradual improvement in the learned routing policy. The early training phase is characterized by high variance in reward, reflecting exploratory behavior and instability in the initial policy. Although oscillations persist throughout training due to environmental stochasticity and continued exploration, the reward trajectory stabilizes within a higher-performance regime after approximately 400–500 episodes, suggesting practical convergence of the learning process. These results demonstrate that PPO is capable of learning effective routing strategies in fault-prone torus networks while maintaining stable performance under dynamic traffic and fault conditions.

Taken together, these results demonstrate that RL-based fault-adaptive routing consistently outperforms adaptive routing across all tested scenarios. Its ability to maintain high throughput under heavy load, preserve delivery ratios under substantial fault densities, and sustain a robust composite fault-adaptive score provides strong evidence that reinforcement learning—when paired with a topologically rich structure like the torus—offers a scalable and resilient alternative to classical routing. The experimental findings underscore the central strength of the RL approach: rather than relying on rigid rules, the agent adapts fluidly to the dynamic network environment, continuously optimizing routing decisions to improve overall performance in the face of failures and congestion.

\section{Conclusion}
\label{section:conclusion}
We evaluated RL-based reinforcement learning for adaptive routing in 2D torus Networks-on-Chip under node faults. RL consistently outperformed adaptive routing in throughput, packet delivery ratio, and fault avoidance, maintaining high performance even during significant portion of node failures in the network. It exploited non-minimal detours and wrap-around paths that adaptive schemes could not. These results demonstrate that RL routing actively learns congestion- and fault-aware strategies, leveraging the torus topology’s multi-path diversity. Adaptive routing, in contrast, fails once minimal paths are blocked. Overall, deep reinforcement learning offers a robust and resilient approach for high-performance NoC routing. 

Future work includes scaling the RL to larger or irregular NoC topologies, exploring multi-agent RL for improved router coordination, and evaluating hardware feasibility with lightweight or quantized models. Hybrid RL-heuristic approaches could combine safety guarantees with adaptivity, while online or continual learning would enable long-term adaptation to evolving workloads and faults. These directions aim to advance RL-based routing from simulation toward practical, deployable solutions for next-generation many-core systems.

\bibliographystyle{elsarticle-num}
\bibliography{ref}

\end{document}